\documentclass[aps,pre,reprint,superscriptaddress]{revtex4-1}

\usepackage{amssymb,amsmath,bm}
\usepackage{graphicx}
\usepackage{enumerate}
\usepackage[utf8]{inputenc}
\usepackage{cancel}
\usepackage{color}
\usepackage[colorlinks,linkcolor=blue,urlcolor=blue,citecolor=blue]{hyperref}
\usepackage{float}
\usepackage{soul}

\begin{document}

\def\be{\begin{equation}}
\def\ee{\end{equation}}
\def\bea{\begin{eqnarray}}
\def\eea{\end{eqnarray}}
\def\l{\label}

\newcommand{\eref}[1]{Eq.~(\ref{#1})}%
\newcommand{\Eref}[1]{Equation~(\ref{#1})}%
\newcommand{\fref}[1]{Fig.~\ref{#1}} %
\newcommand{\Fref}[1]{Figure~\ref{#1}}%
\newcommand{\sref}[1]{Sec.~\ref{#1}}%
\newcommand{\Sref}[1]{Section~\ref{#1}}%
\newcommand{\aref}[1]{Appendix~\ref{#1}}%
\newcommand{\sgn}[1]{\mathrm{sgn}({#1})}%
\newcommand{\erfc}{\mathrm{erfc}}%
\newcommand{\erf}{\mathrm{erf}}%

\title{Asymmetric Stochastic Resetting: Modeling Catastrophic Events}

\author{Carlos A. Plata}
\affiliation{Dipartimento di Fisica `G. Galilei', INFN, Universit\`a di Padova, Via Marzolo 8, 35131 Padova, Italy}
\author{Deepak Gupta}
\affiliation{Dipartimento di Fisica `G. Galilei', INFN, Universit\`a di Padova, Via Marzolo 8, 35131 Padova, Italy}
\author{Sandro Azaele}
\address{Dipartimento di Fisica `G. Galilei', INFN, Universit\`a di Padova, Via Marzolo 8, 35131 Padova, Italy}
\date{\today}

\begin{abstract}
In the classical stochastic resetting problem, a particle, moving according to some stochastic dynamics, undergoes random interruptions that bring it to a selected domain, and then, the process recommences. Hitherto, the resetting mechanism has been introduced as a symmetric reset about the preferred location. However, in nature, there are several instances where a system can only reset from certain directions, e.g., catastrophic events. Motivated by this, we consider a continuous stochastic process on the positive real line. The process is interrupted at random times occurring at a constant rate, and then, the former relocates to a value only if the current one exceeds a threshold; otherwise, it follows the trajectory defined by the underlying process without resetting. An approach to obtain the exact non-equilibrium steady state of such systems and the mean first passage time to reach the origin is presented. Furthermore, we obtain the explicit solutions for two different model systems. Some of the classical results found in symmetric resetting such as the existence of an optimal resetting, are strongly modified. Finally, numerical simulations have been performed to verify the analytical findings, showing an excellent agreement.
\end{abstract}

\maketitle

\section{Introduction}
Ecosystems regularly undergo either environmental or anthropogenic disturbances which alter the number of species as well as the size of their populations. Natural disasters or catastrophes, such as droughts, fires,  epidemics or invasions may cause major declines. In the aftermath of these, depleted populations have to recover from low population sizes with an increased risk of extinction \cite{Gerber2001,Assaf2009}. Similarly, financial crashes affect gross domestic product, asset prices, consumptions and investments, and therefore, strongly modify typical business cycles \cite{Mendoza2010,Sornette2017}. 

These two examples show that, besides being rare and extreme, such events are not followed by episodes of comparable large increases in the corresponding variables. Explaining abrupt crashes is challenging, especially when trying to find a general tool applicable to a large class of stochastic models. Indeed, these crises have the potential to alter the temporal dynamics of state variables as well as the steady state properties of the system. 

In this work, we introduce a toy framework which can be applied to a large class of stochastic processes and can account for abrupt changes in some state variable. It deals with the effects of sudden drops by introducing random resetting events to a non-vanishing value within a diffusive stochastic process.

As it stands today, stochastic resetting was originally introduced in the context of search processes \cite{Evans_2011, Evans_2011b, kusmierz_2014, Chechkin_2018, Belan_2018, Evans_2020}. Remarkably,  its foundation has brought also a collection of appealing results that include the non-equilibrium steady state \cite{mendez_2016, Pal_2016b,underdamped,Restart-KPZ, transport1,Pal-potential,invariance,invariance2,SEP}, optimization of the mean first passage time \cite{Reuveni_2016, Pal_2016,branching}, and fluctuation theorems \cite{Fuchs_2016, Pal_2017, restart_conc17, Busiello_2020, Gupta_2020}.

Stochastic resetting has been used in a plethora of applications \cite{Evans_2020}. In the context of population dynamics, resetting is known as catastrophe and mimics  the effects of natural disasters in the ecosystem. Indeed, some of the primordial notions of stochastic resetting for the modeling of catastrophic events can be found in the literature \cite{levikson_1977, pakes_1978, brockwell_1982, brockwell_1985, kyriakidis_1994, Pakes_1997, economou_2003}. However, these models have been usually described through a dynamics based on jump processes in which resetting is added. The main goal of this paper is to apply a comprehensive theoretical framework provided by Markov processes with reset to population dynamics described through diffusion processes.

Mimicking the perturbation produced by a natural disaster or a sudden financial drop using stochastic resetting forces us to re-define the assumptions of the relocations. More specifically, the reset events have to be asymmetric, i.e., albeit the population size (or the particle position) may plummet owing to a catastrophic event, it is nevertheless impossible that an offsetting positive increment of the variable occurs owing to another similar event. 

Motivated by this, we present an approach to a general problem of asymmetric stochastic resetting in diffusive processes. We apply it to two paradigmatic examples which exemplify the main features and consequences of such asymmetry. Herein, we tackle the following two relevant questions: \textit{i)} What is the hallmark of such a resetting mechanism at stationarity? In other words, how is the non-equilibrium stationary state modified due to resetting? \textit{ii)} How does the mean lifetime of a population change under asymmetric stochastic resetting? 

The remaining of the paper is organized as follows. In Sec.~\ref{model}, we discuss the basis of our model to mimic population dynamics involving catastrophic events. The non-equilibrium steady state and the mean first passage time, respectively, are discussed in Secs. \ref{secness} and \ref{secmfpt}. Therein, we introduce a general formalism that afterwards particularized to two specific situations of interest. Finally, we present the main conclusion of our work in Sec. \ref{seccons}. Some technical details and lengthier auxiliary calculations are shown in Appendices.

\section{Model}
\label{model}
We approximate the evolution of the population size, i.e., the number of individuals, of a given species by a continuous-state stochastic process defined on the positive real line. Starting with a positive population size, at later times the number of individuals, $ x $, is governed by the following Langevin dynamics,
\begin{align}
\dfrac{dx}{dt}= A(x)+\sqrt{2B(x)} \eta(t),
\label{dyn-1}
\end{align}
where $A(x)$ and $B(x)$ ($A(0) > 0$ and $B(0) = 0$ in population dynamics), respectively, are the state-dependent drift and diffusion terms. Also, $\eta(t)$ is a Gaussian white noise with zero mean and delta correlated, i.e., $\langle \eta(t) \rangle=0$ and $\langle \eta(t) \eta(t^\prime) \rangle=\delta(t-t^\prime)$. The above stochastic differential equation has to be interpreted according to the It\^o scheme \cite{vK}. Diffusive models have been proven to be very useful to capture emergent patterns in population dynamics \cite{hubbell2001unified,azaele2006,Volkov2007,azaele2016neutral}. Remarkably,  Eq.~\eqref{dyn-1} is very general, allowing the study of generalized models involving heterogeneous diffusion, which are processes of great interest in the field of anomalous diffusion \cite{ref-h-1,ref-h-2,ref-h-3,ref-h-4,ref-h-5}.

\begin{figure}
  \begin{center}
    \includegraphics[width=0.3\textwidth]{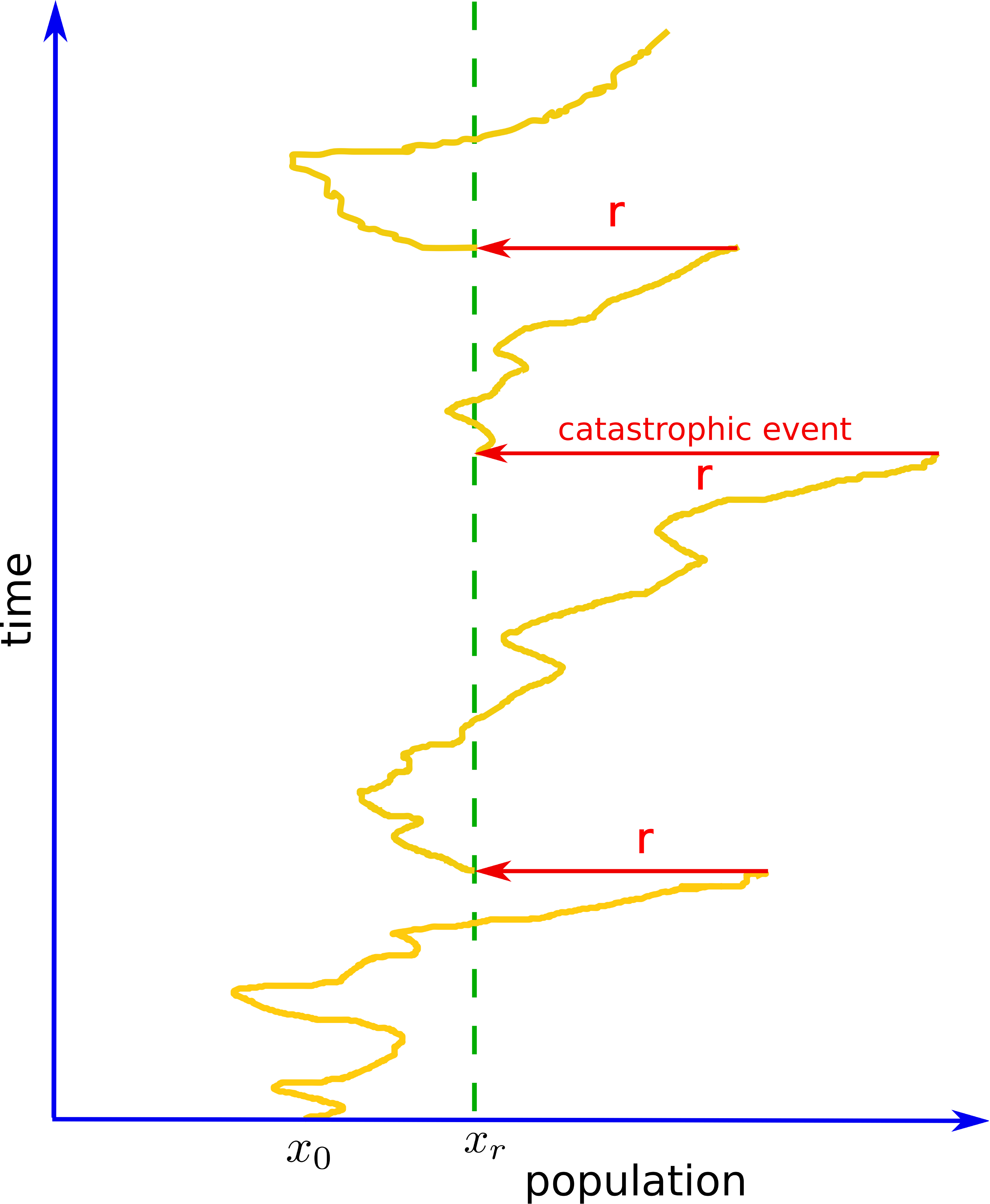}
    \caption{Sketch of asymmetric stochastic resetting process. The system evolves over time (zigzag curve) with interrupting events (horizontal arrows) which bring the former to a certain value $x_r>0$ with rate $r(x)=r \Theta(x-x_r)$.}
    \label{fig:scheme}
  \end{center}
\end{figure}

In addition to the dynamics described by Eq.~\eqref{dyn-1}, we assume that there is a stochastic resetting to a constant value $x_r>0$. The resetting events occur at random times with a constant rate $r$, but only if the population size is above the resetting threshold $x_r$. The schematic representation of such a composed process is shown in Fig. \ref{fig:scheme}.

Of course, different choices of $A(x)$ and $B(x)$ lead to completely different stochastic models, with very different physical properties. Nevertheless, it is possible to study some relevant features of the process with a unified approach, which we develop here. Later on, we will look into two more specific cases, which have important applications: (I) pure homogeneous diffusion; (II) simple population dynamics with demographic stochasticity.

The resetting mechanism introduced above is a particular but relevant case of resetting. It corresponds to regular stochastic resetting with a state-dependent rate $r(x)=r \Theta(x-x_r)$, where $\Theta(\cdot)$ is the Heaviside function that guarantees that resetting only occurs when population size is larger than $x_r$, in contrast to standard symmetric resetting mechanism. This is another appealing aspect of our modeling since the formulation of state-dependent resetting rates has been already introduced \cite{Evans_2011b, roldan_2017, pinsky_2020} but the applications have been, to the best of our knowledge, quite scarce.

The dynamics of the propagator $p(x,t|x_0)$, which is the probability of reaching the state $x$ at time $t$ departing from initial state $x_0$ at time zero, is captured by the master equation:
\begin{align}
\label{eq:r-ev}
\dfrac{\partial p(x,t|x_0)}{\partial t} =&  -\dfrac{\partial J(x,t|x_0)}{\partial x}- r\Theta (x-x_r)p(x,t|x_0)\nonumber\\ &+ r \delta (x-x_r) \int_{x_r}^{\infty} dy~ p(y,t|x_0),
\end{align}
where $J(x,t|x_0):=A(x) p(x,t|x_0)-\partial_x [B(x)p(x,t|x_0)]$ is the probability flux that stems from the resetting-free dynamics in Eq.~\eqref{dyn-1}. 
The second term on the right hand side corresponds to the loss rate of the probability from $ x $
due to resetting, while the third term represents the corresponding gain rate of the probability at $ x = x_r $ coming from the resetting of all positions larger than $ x_r $. Note that the resetting term into Eq.~\eqref{eq:r-ev} is a particularization of the general term for state-dependent resetting rate, firstly introduced in \cite{Evans_2011b}, for our specific choice of asymmetric resetting.

\section{Non-equilibrium stationary state}
\label{secness}
As stated in the introduction, the non-equilibrium
steady state for the symmetric resetting has been already
studied in the literature \cite{
Evans_2020,Pal_2016b,mendez_2016,underdamped,Restart-KPZ,transport1,Pal-potential,invariance,invariance2,SEP}. However, in this paper, we study the models where the resetting is asymmetric with respect to its resetting location. Herein, we focus on the study of the non-equilibrium stationary state of Eq.~\eqref{eq:r-ev}, $p_{ss}(x)$, subject to reflecting boundary conditions at $ x=0 $. 
We can obtain $p_{ss}(x)$  by setting the left hand side of the Eq.~\eqref{eq:r-ev} to zero, and solving for the distribution. Since we have to deal with a discontinuity in the equation \eqref{eq:r-ev}, it is handy to define $P_L(x)$ and $P_R(x)$ as the stationary solutions to the left and to the right of $x_r$, respectively. Therefore, the corresponding fluxes $J_L(x)$ and $J_R(x)$ obey the following equations
\begin{subequations}
\label{eq:r-ev-LR}
\begin{align}
\label{eq:r-ev-L}
 \partial_x J_L(x)&=0,  &0<x< x_r, \\
\label{eq:r-ev-R}
 \partial_x J_R(x)&= - r  P_R(x),  &x>x_r.
\end{align}
\end{subequations}
Note that, in our problem, it is convenient to study the current contributions explicitly in this way. 
These equations have to be complemented with the boundary conditions 
\begin{subequations}
\label{eq:r-bc}
\begin{align}
\label{eq:r-bc-L}
J_L(0)&=0,
\\
\label{eq:r-bc-R}
\lim_{x \to \infty} J_R(x)&=0,
\end{align}
\end{subequations}
and the matching conditions
\begin{subequations}
\label{eq:r-mc}
\begin{align}
\label{eq:r-mc1}
P_R(x_r)&=P_L(x_r),
\\
\label{eq:r-mc2}
J_R(x_r)&=J_L(x_r)+r \int_{x_r}^{\infty} dx \, P_R(x).
\end{align}
\end{subequations}
Eq.~\eqref{eq:r-mc1} is the continuity condition for our solution, whereas the kink condition in Eq.~\eqref{eq:r-mc2} is obtained by integrating Eq.~\eqref{eq:r-ev} from $x_r-\epsilon$ to $x_r+\epsilon$ and then taking the limit $\epsilon \to 0^+$.

Since there is no probability leakage from the boundaries, the normalization is preserved over the whole evolution,
\begin{equation}
\label{eq:r-norm}
\int_{0}^{x_r} dx \, P_L(x,t)+\int_{x_r}^{\infty} dx \, P_R(x)=1.
\end{equation}
It could seem that we have an excess of conditions, since we have two second order ODEs \eqref{eq:r-ev-LR}, and five conditions to fulfill, i.e., Eqs. \eqref{eq:r-bc}, \eqref{eq:r-mc} and \eqref{eq:r-norm}. This apparent paradox is resolved when studying carefully the kink condition \eqref{eq:r-mc2}. Integrating Eq.~\eqref{eq:r-ev-R} from $x_r$ to $\infty$, using the boundary conditions \eqref{eq:r-bc-R}, and taking into account that $J_L(x)=0$, one obtains the matching condition \eqref{eq:r-mc2}. Thus, the kink condition becomes a trivial identity that always holds. 

Let us first focus on the region $0<x<x_r$. We have to solve Eq.~\eqref{eq:r-ev-L} with the reflecting boundary condition defined in \eqref{eq:r-bc-L}.  
This is a first order linear ODE for $P_L(x)$ whose solution is determined up to an arbitrary constant $\mathcal{N}_1$:
\begin{equation}
\label{sol-L}
P_L(x)=  f_L(\mathcal{N}_1, x),
\end{equation} 
where
\begin{equation}
f_L(\mathcal{N}_1, x)= \frac{\mathcal{N}_1}{B(x)} \exp \left[ \int^x \!\!dy \frac{A(y)}{B(y)}\right],
\end{equation}
that is, the equilibrium solution \cite{vK} of the stochastic model without resetting.

When $ x $ is larger than $ x_r $, we solve Eq.~\eqref{eq:r-ev-R} with a reflecting boundary at infinity, i.e., Eq.~\eqref{eq:r-bc-R}. Thus, the general solution is given by 
\begin{equation}
\label{sol-R}
P_R(x)=  f_R(\mathcal{N}_2, x),
\end{equation}
determined up to another arbitrary constant $\mathcal{N}_2$.
The constants $\mathcal{N}_1$ and $\mathcal{N}_2$ can be found using conditions \eqref{eq:r-mc1}  and \eqref{eq:r-norm}. 

Hitherto, we have outlined a procedure to obtain the solution for arbitrary smooth functions $A(x)$ and $B(x)$. Clearly, the choice of a specific stochastic model is crucial and could lead to computational difficulties in the determination of an explicit solution, especially in the calculation of $f_R$ \eqref{sol-R}. In order to appreciate analogies and differences with processes with symmetric resetting, in the following, we have considered two prototypical cases of stochastic processes submitted to asymmetric resetting. As well as being of intrinsic theoretical importance, they are also relevant in applications. 

In the first case (I), we consider a particle which undergoes pure diffusion with diffusive constant $D$ on the real
positive line. When hitting the origin, it bounces back to the positive domain, whereas when (and only when)
its position is larger than $ x_r $, it is re-located at $ x = x_r $ at random times with a constant rate $ r $. One obtains the stationary distribution (see Appendix \ref{sec:app_pss} for details):
\begin{align}
p_{ss}(x)=
\begin{cases}
\dfrac{1}{x_r+ \sqrt{D/r}}~& \text{for } 0 \leq x \leq x_r,\\
\dfrac{\exp \left[-\sqrt{r/D}(x-x_r)\right]}{x_r+ \sqrt{D/r}}~&  \text{for } x>x_r.
\end{cases}
\label{pss-I}
\end{align} 
Note that the probability of finding the particle at positions smaller than $ x_r $ is uniform, whereas there is an exponential decay for $ x > x_r $. The exponential decay is the hallmark of standard diffusion \cite{Evans_2011} with symmetric resetting, whereas, in the region without resetting, we recover the uniform solution.

In the second case (II), we consider an ecological model defined by $ A(x) = b - x $ and $ B(x) = x $. The details of its derivation are presented in Appendix \ref{sec:app_dimensionless}. The drift term accounts for immigration and net death rate of individuals in a certain region. Instead, $ B(x) $ is linear on the population size, because the model assumes that the source of stochasticity is only due to individual random births and deaths. This model 	 has been used to explain some macro-ecological patterns in species-rich ecosystems \cite{azaele2016neutral,Peruzzo2020}. In this case the asymmetric resetting describes how the population size plummets to a smaller size in the aftermath of environmental catastrophic events. The solution for the stationary distribution of this ecological model reads 
\begin{align}
p_{ss}(x)=
\begin{cases}
\mathcal{N} x^{-1+b}e^{-x}~& \text{for } 0 \leq x \leq x_r,\\
\mathcal{N} x^{-1+b}e^{-x} \dfrac{U(r,b,x)}{U(r,b,x_r)}~&  \text{for } x>x_r,
\end{cases}
\label{pss-II}
\end{align} 
where $\mathcal{N}$ is a normalization constant (see Appendix \ref{sec:app_pss} for further details) and $U(\alpha,\beta,x)$ is the confluent hypergeometric function of the second kind \cite{Lebedev1972}. Remarkably, $ p_{ss}(x) $ is a very well known function in theoretical ecology, which is used to quantify the total number of species with a given number of individuals within some spatial region. In diffusive models of population dynamics, this empirical pattern is usually well approximated by a gamma distribution  \cite{hubbell2001unified,azaele2006,Volkov2007,azaele2016neutral} when there is no resetting.
 \begin{figure}
  \begin{center}
    \includegraphics[width=4cm]{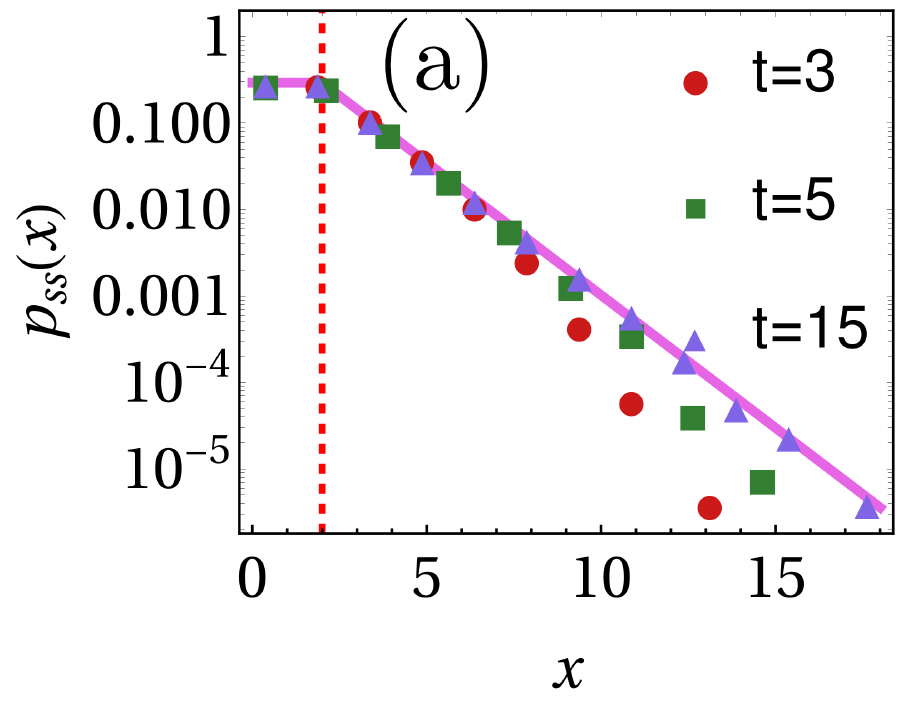}~~~
    \includegraphics[width=4cm]{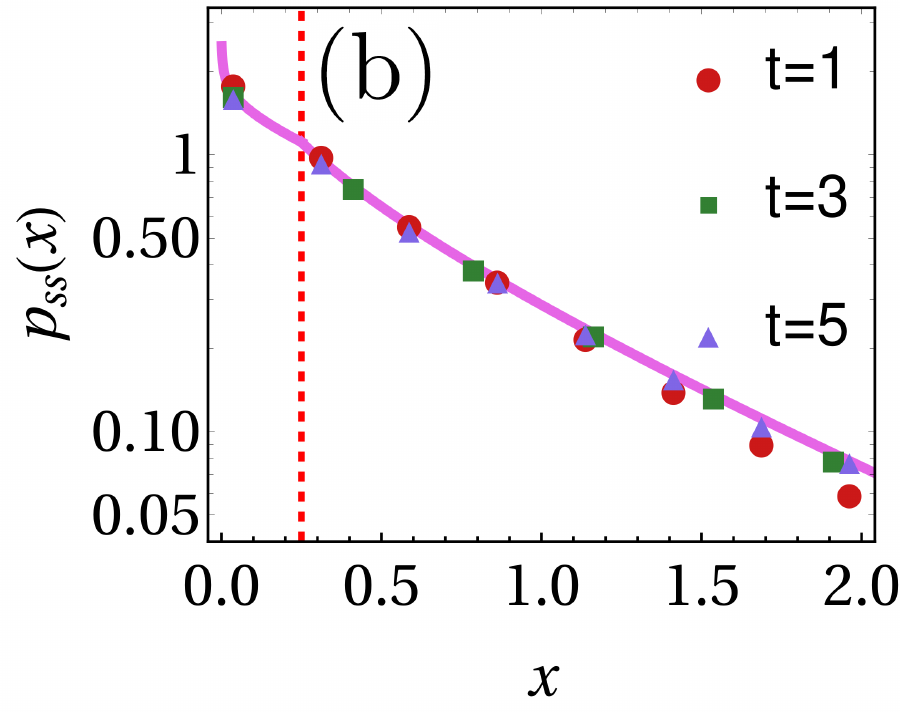}
    \caption{Relaxation to the steady state distribution $p_{ss}(x)$. Panel (a): (I) pure diffusion. Panel (b): (II) ecological model. The solid curve stands for the stationary theoretical prediction while circles, squares, and triangles are obtained from the numerical simulation at three different times. The parameters for panel (a) are $x_0=x_r=2$, $r=0.5$, and $D=1$; and for panel (b) $b=0.9$, $x_0=x_r=0.25$, and $r=0.5$. In each case, the vertical dashed line corresponds to the resetting location $x_r$.}
    \label{fig:ss-dist}
  \end{center}
\end{figure}

In Fig.~\ref{fig:ss-dist}, we compare the theoretical prediction (solid curve) of the steady state distribution $p_{ss}(x)$ given in Eqs.~\eqref{pss-I} and \eqref{pss-II} with the distribution obtained by numerical simulations (circles, squares, and triangles) at three different times. 
Herein, we have taken the initial condition equal to $ x_r $, but this has no effect on the final stationary state. Notice that, as the observation time increases, the difference between theory and finite time simulations decreases, up to becoming negligible within the plotted range, since simulations have reached the stationary regime.

\section{Mean first passage time}
\label{secmfpt}
To study the mean first passage time (MFPT) to reach $ x = 0 $, we have to assume that the origin of the real axis is an absorbing boundary. If the probability to hit that boundary is one as $t \to \infty$, then the equation for the MFPT departing from $ x $, $\tau(x)$, is
\begin{align}
\label{eq:tau-ev-m}
-1=&A(x) \partial_{x}\tau(x)+B(x) \partial_x^2 \tau(x) \nonumber \\ &+r \Theta(x-x_r)[\tau(x_r)-\tau(x)],
\end{align}
A comprehensive derivation of the above equation based on the backward master equation \eqref{eq:r-ev} is reserved in Appendix \ref{sec:app_eqMFPT}. This equation has to be complemented with the boundary conditions
\begin{subequations}
\label{eq:tau-bc}
\begin{align}
\label{eq:tau-bc1}
&\tau(0)=0,
\\
\label{eq:tau-bc2-m}
\lim_{x \to \infty} &\tau(x) ~ \text{is  finite}.
\end{align}
\end{subequations}
Note that the presence of resetting entails a finite MFPT as $x \to \infty $, since the reset connects any value of $x>x_r$ with $x_r$.

In order to find the solution of Eq.~\eqref{eq:tau-ev-m}, we follow a strategy similar to before: solving the equation to both sides of $x_r$ separately and then imposing the proper boundary and matching conditions. In the following, we present the solutions for the two cases of interest we have introduced previously. The detailed derivation is relegated to the Appendix \ref{sec:app_MFPT}.

\begin{figure}
  \begin{center}
    \includegraphics[width=4cm]{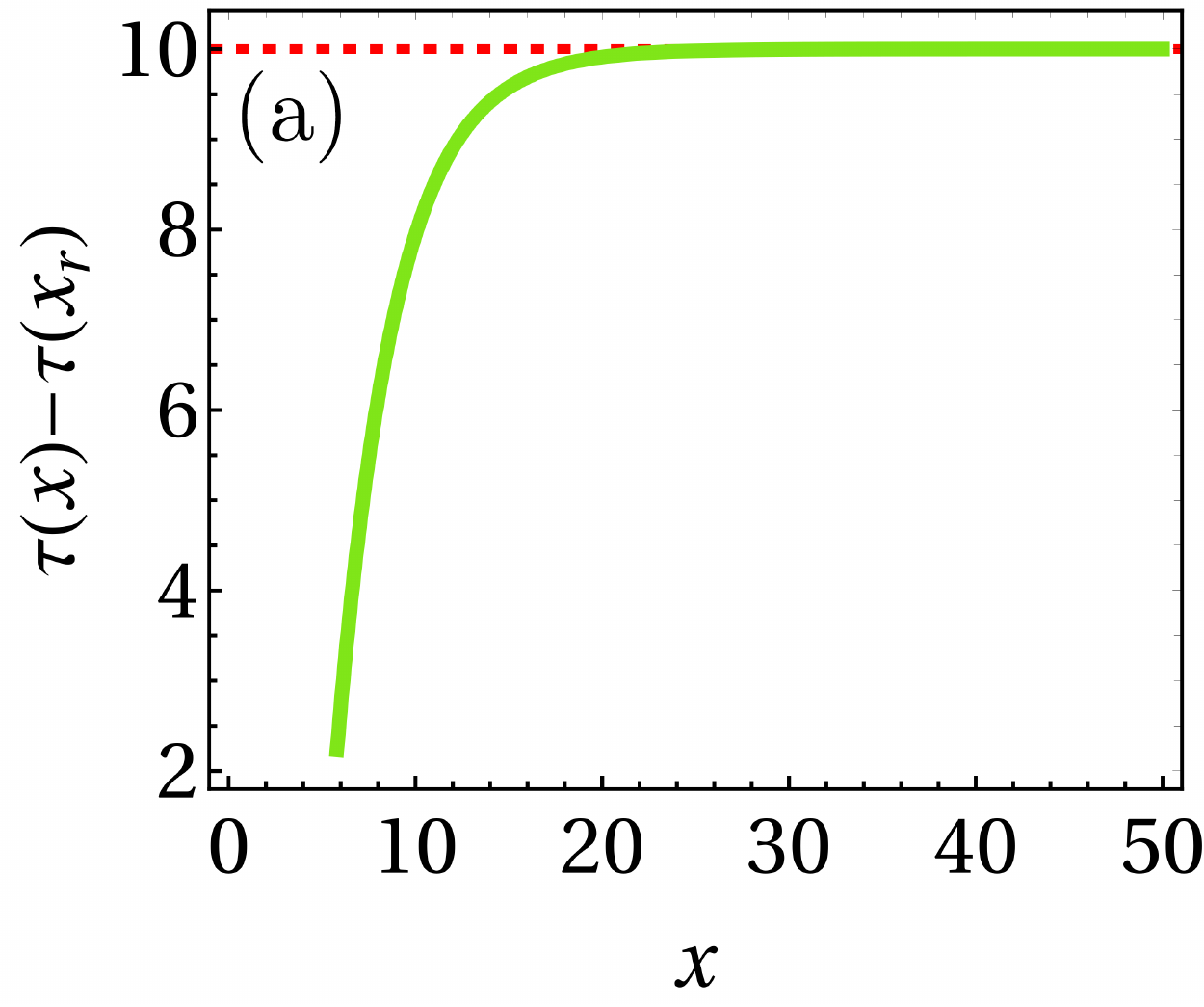}~~~
    \includegraphics[width=4cm]{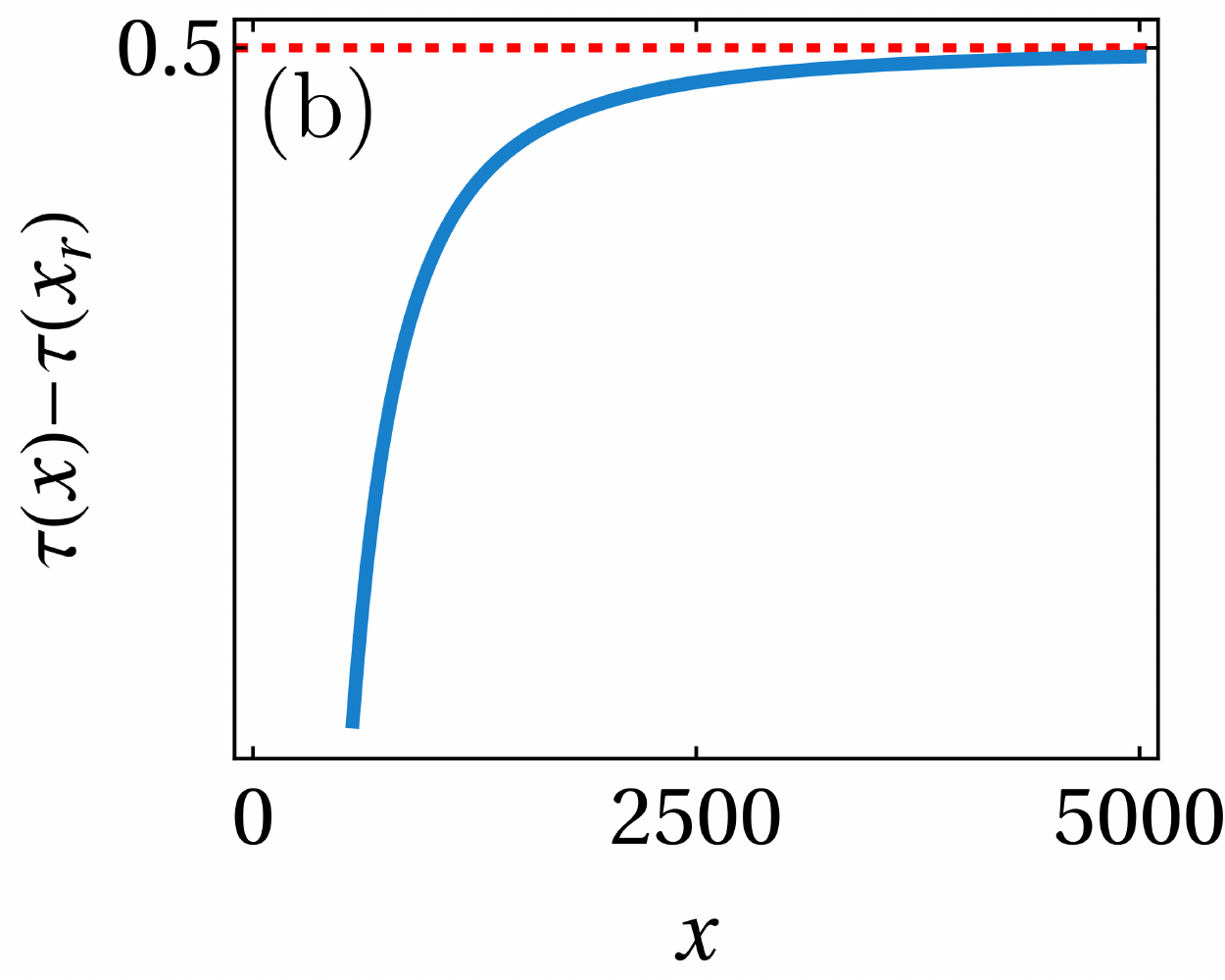}
    \caption{Asymptotic of $[\tau(x)-\tau(x_r)]$ with respect to $x$. Clearly, we can see that $[\tau(x)-\tau(x_r)]$ approaches $r^{-1}$ (horizontal red dashed line) as $x\to\infty$. The parameters used in the above plots for diffusion model (panel (a)) are $D=1,~x_r=5,~r=0.1$ and for ecological one  (panel (b)) are $x_r=1.0,~b=0.8,~r=2.0$.}
    \label{large-x}
  \end{center}
\end{figure}

In the case of pure diffusion the MFPT reads
\begin{align}
\label{mfpt-diff-process-m}
\tau(x)=\begin{cases}
- \dfrac{x^2}{2D} + x\left(\dfrac{x_r}{D}+ \dfrac{1}{\sqrt{rD}} \right)  &0 \leq x \leq x_r,\\
\\
\dfrac{1-e^{-\sqrt{r/D}(x-x_r)}}{r}+\dfrac{x_r^2}{2D} + \dfrac{x_r}{\sqrt{rD}}& x > x_r.
\end{cases}
\end{align}
On the other hand, the mean lifetime in the ecological case equals to 
\begin{equation}
\label{mfpt-db-process-m}
\tau(x)=\tau_L(x) \Theta(x_r-x)+\tau_R(x)\Theta(x-x_r),
\end{equation}
where
\begin{subequations}
\begin{align}
\tau_L(x)=& \int_0^x  \, dy \, y^{-b}e^y \bigg[ \Gamma(b,y) - \Gamma (b,x_r) \nonumber
\\
&+ \frac{U(1+r,1+b,x_r)}{U(r,b,x_r)} x_r^b e^{-x_r} \bigg], \label{eq:tauL} \\
\tau_R(x)= &\tau_L(x_r) + \frac{1}{r} \left[ 1- \frac{U(r,b,x)}{U(r,b,x_r)}\right].
\end{align}
\end{subequations}
Note that  $ \lim_{x \to \infty}\tau(x)-\tau(x_r)=1/r$ in both cases, as shown in In Fig.~\ref{large-x}. Indeed, this general property can be derived from Eq.~\eqref{eq:tau-ev-m}, when considering Eq.~\eqref{eq:tau-bc2-m} and taking the limit $x\to \infty$. 
\begin{figure}
  \begin{center}
   \includegraphics[width=4cm]{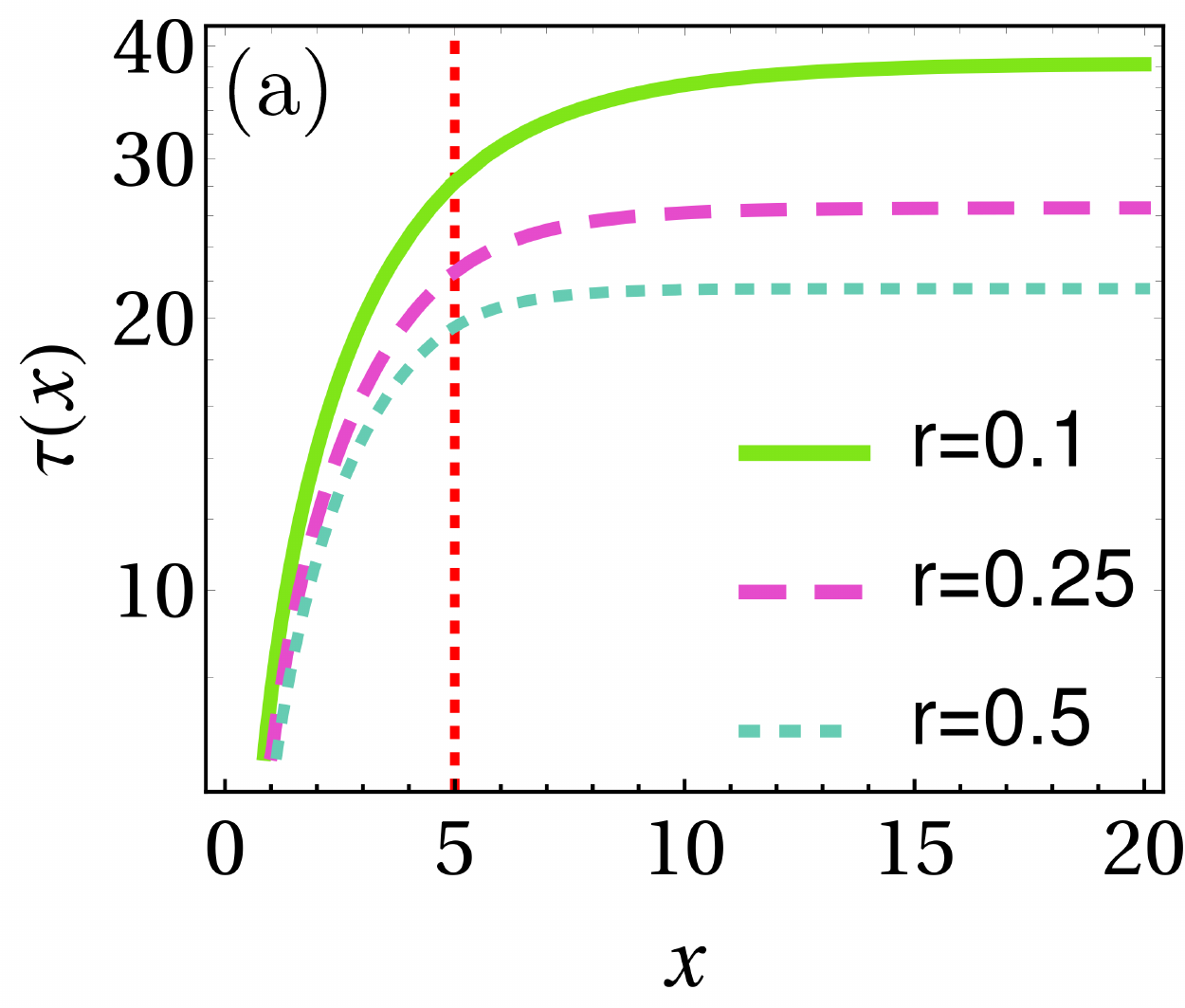}~~~
  \includegraphics[width=4cm]{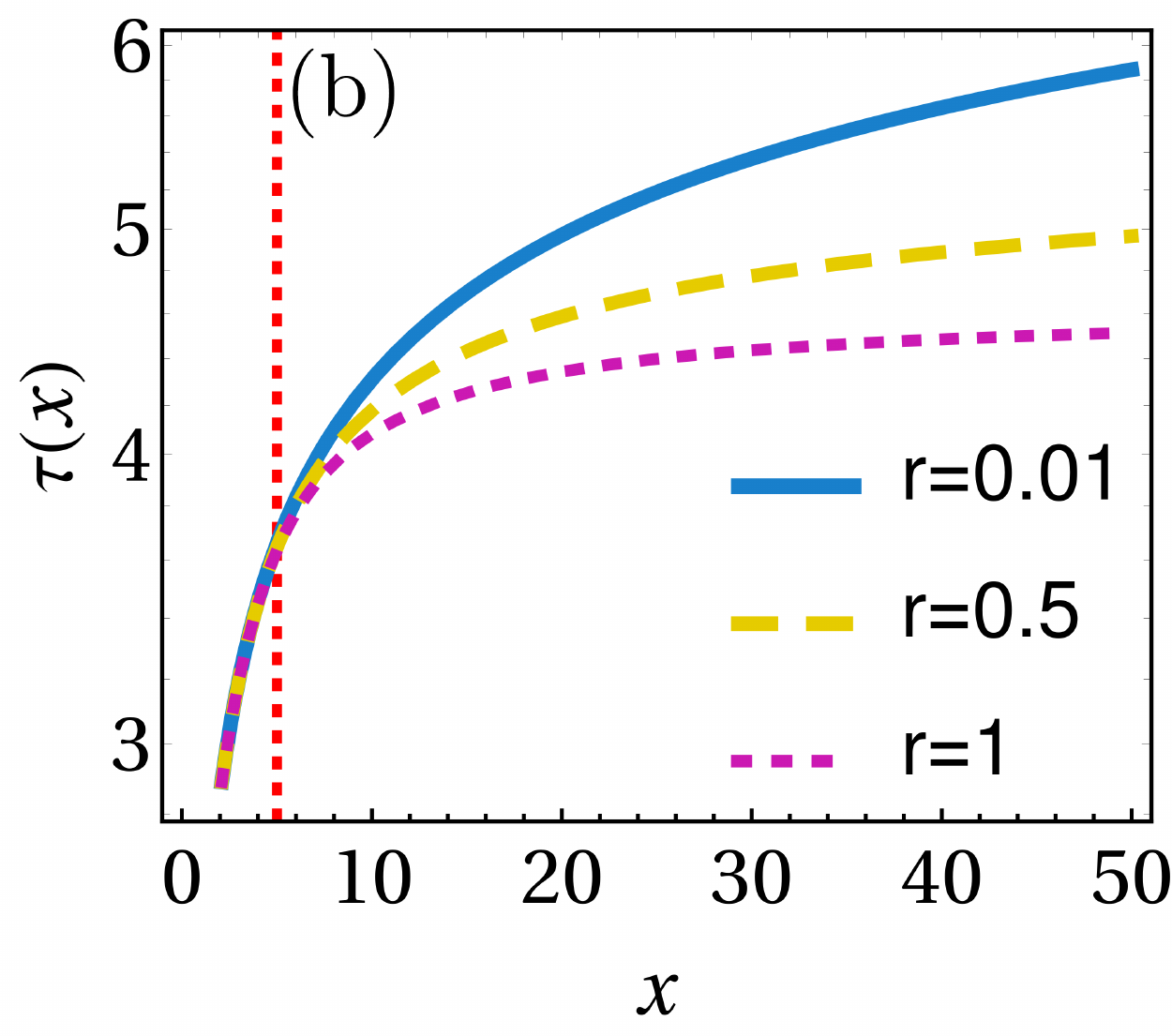}
    \caption{Mean first passage time  $\tau(x)$. Panel (a): (I) pure diffusion. Panel (b): (II) ecological model. It is observed that $\tau(x)$ reaches a constant value for large $x$, and  it increases with the initial location of the system. As it is reasonable, the mean first passage time decreases with resetting rate $r$ for given $x$. The vertical dashed line indicates the resetting location $x_r$. The parameters for panel (a) are $x_r=5$ and $D=1$; and for panel (b) $b=0.5$ and $x_r=5$. In each case, the vertical dashed line corresponds to the resetting location $x_r$.}
    \label{fig:mfpt-diff-bv}
  \end{center}
\end{figure}

\begin{figure}
  \begin{center}
   \includegraphics[width=4cm]{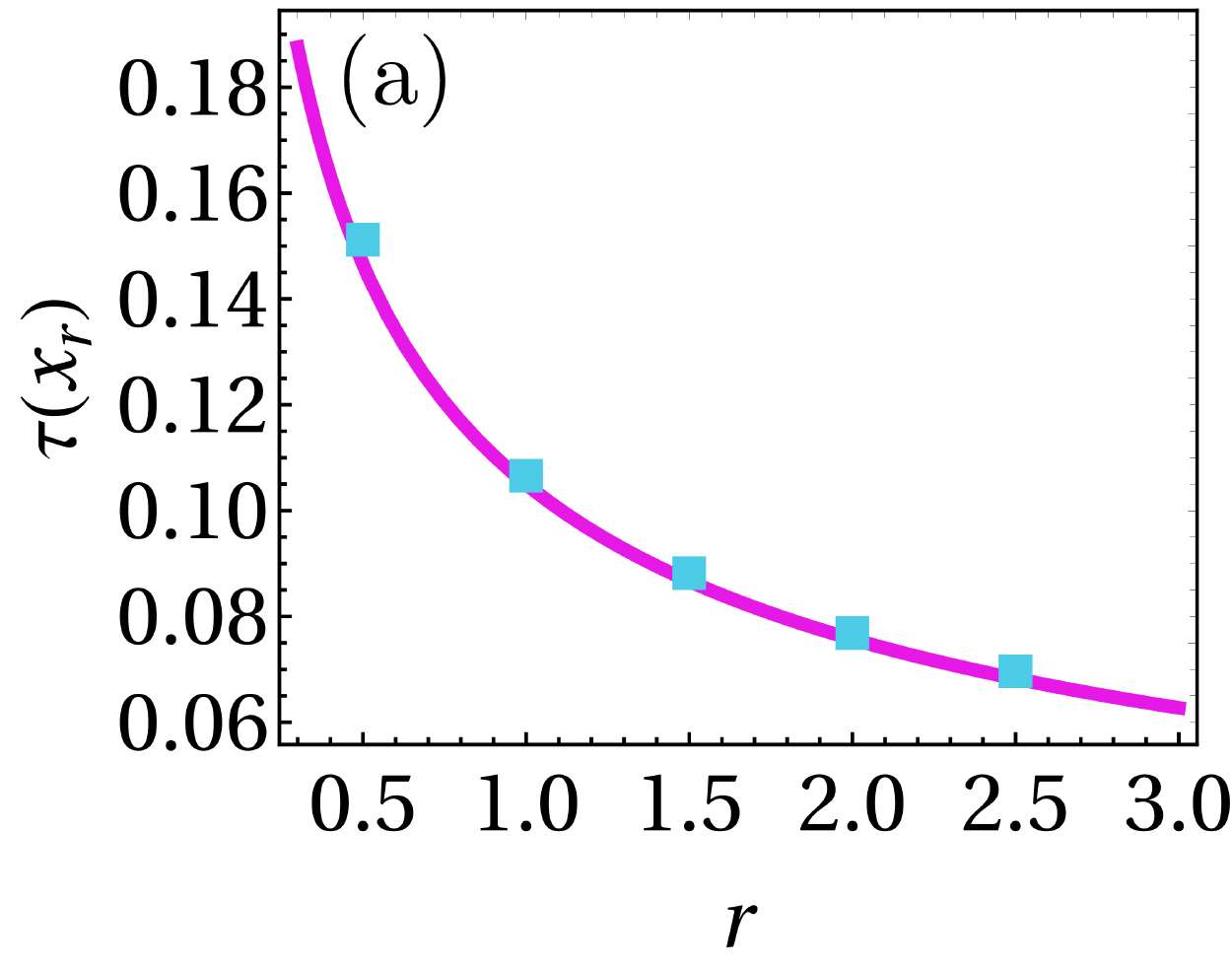}~~~
   \includegraphics[width=4cm]{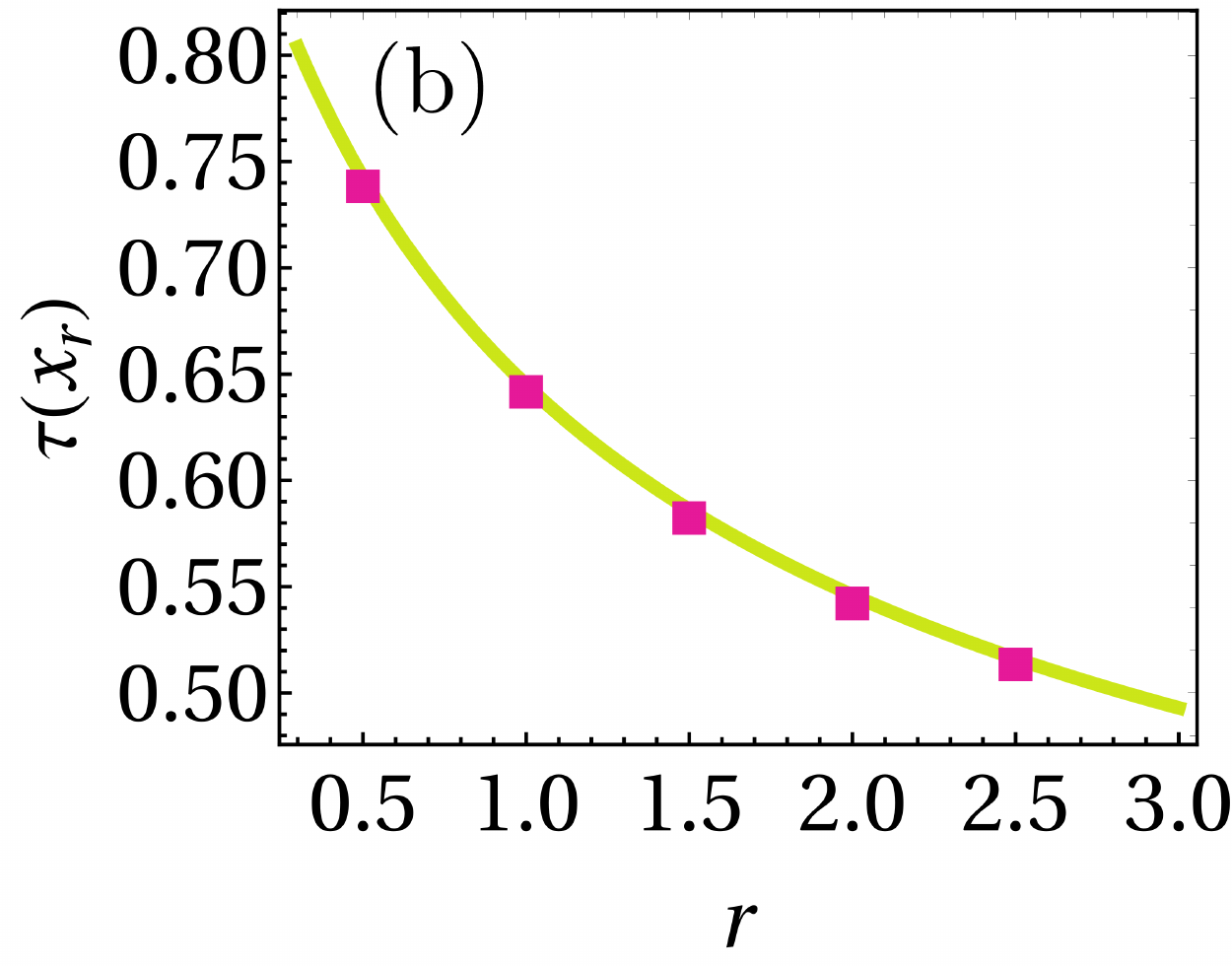}
    \caption{Mean first passage time $\tau(x_r)$ as a function the resetting rate $r$. Panel (a): (I) pure diffusion. Panel (b): (II) ecological model. In both cases, solid curve is the analytical prediction given by Eqs.~\eqref{mfpt-diff-process-m} and \eqref{mfpt-db-process-m} whereas the squares are obtained from numerical simulations. The parameters for the panel (a) are $x_r=x_0=0.1$ and $D=1$; and for the panel (b) $x_r=x_0=0.1$ and $b=0.5$.}
    \label{fig:mfpt}
  \end{center}
\end{figure}

We plot the theoretical MFPT [Eqs.~\eqref{mfpt-diff-process-m} and \eqref{mfpt-db-process-m}] with respect to the initial location $x$ in Fig.~\ref{fig:mfpt-diff-bv} for both cases. For a fixed $r$, it is clear that the MFPT reaches asymptotically a constant value as $x$ increases.  Moreover, we highlight that $\tau(x)$ monotonically decreases as $r$ increases for a fixed $x$ (see Fig.~\ref{fig:mfpt}). This is because the asymmetric resetting brings the system to $x_r$ only when $ x $ is larger than $x_r$. Hence, our results depart from the ones obtained in \cite{Evans_2011b}, since the asymmetry in the resetting makes the dependence monotonic and removes any possibility of an optimal resetting rate, which stemmed from the combined effect of resetting to both sides of $x_r$. Finally, we compare the analytical results of  MFPT [Eqs.~\eqref{mfpt-diff-process-m} and \eqref{mfpt-db-process-m}] with the numerical simulations in Fig.~\ref{fig:mfpt} for both model systems, and they have an excellent agreement.

\begin{figure}
  \begin{center}
    \includegraphics[width=4cm]{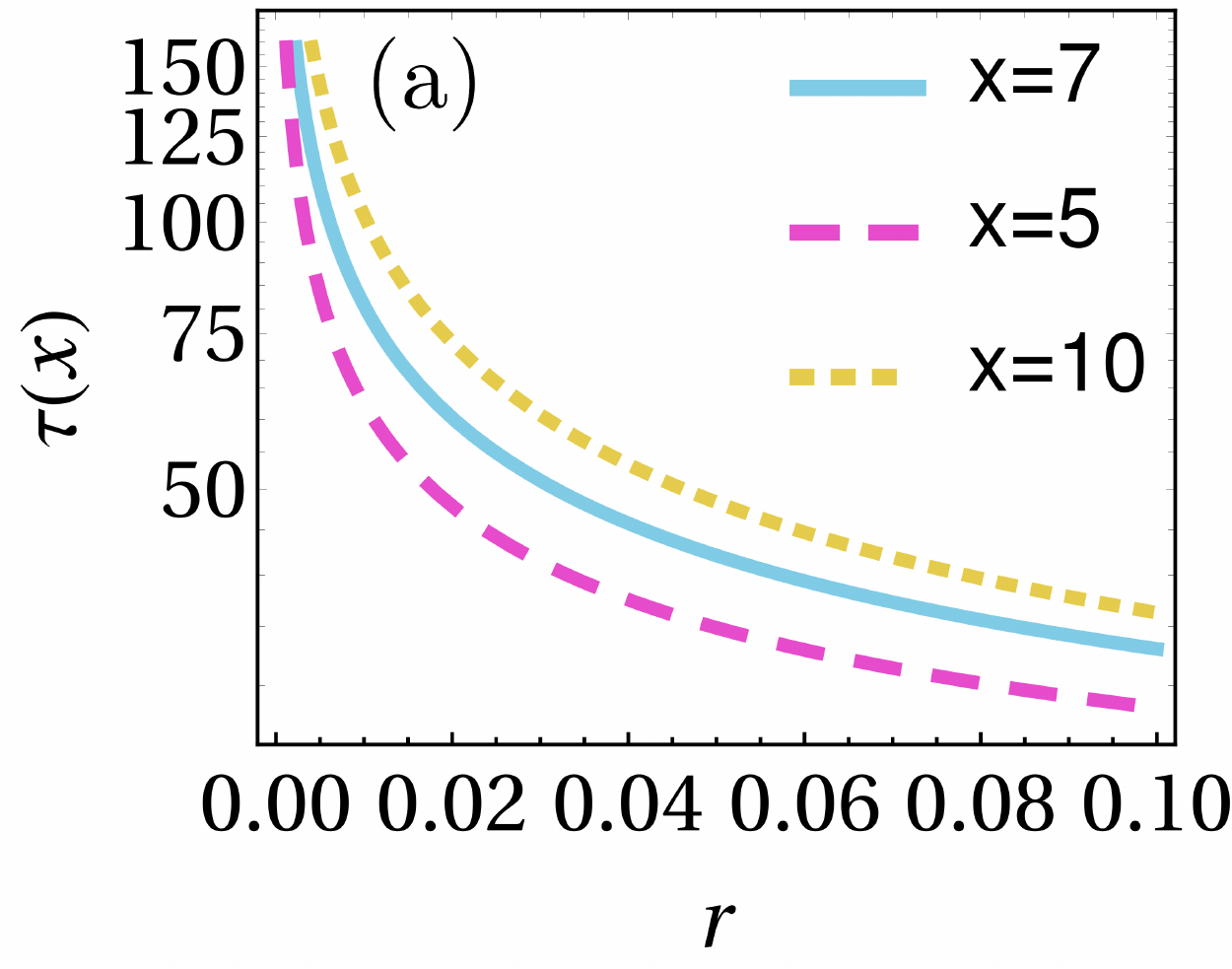}~~~
    \includegraphics[width=4cm]{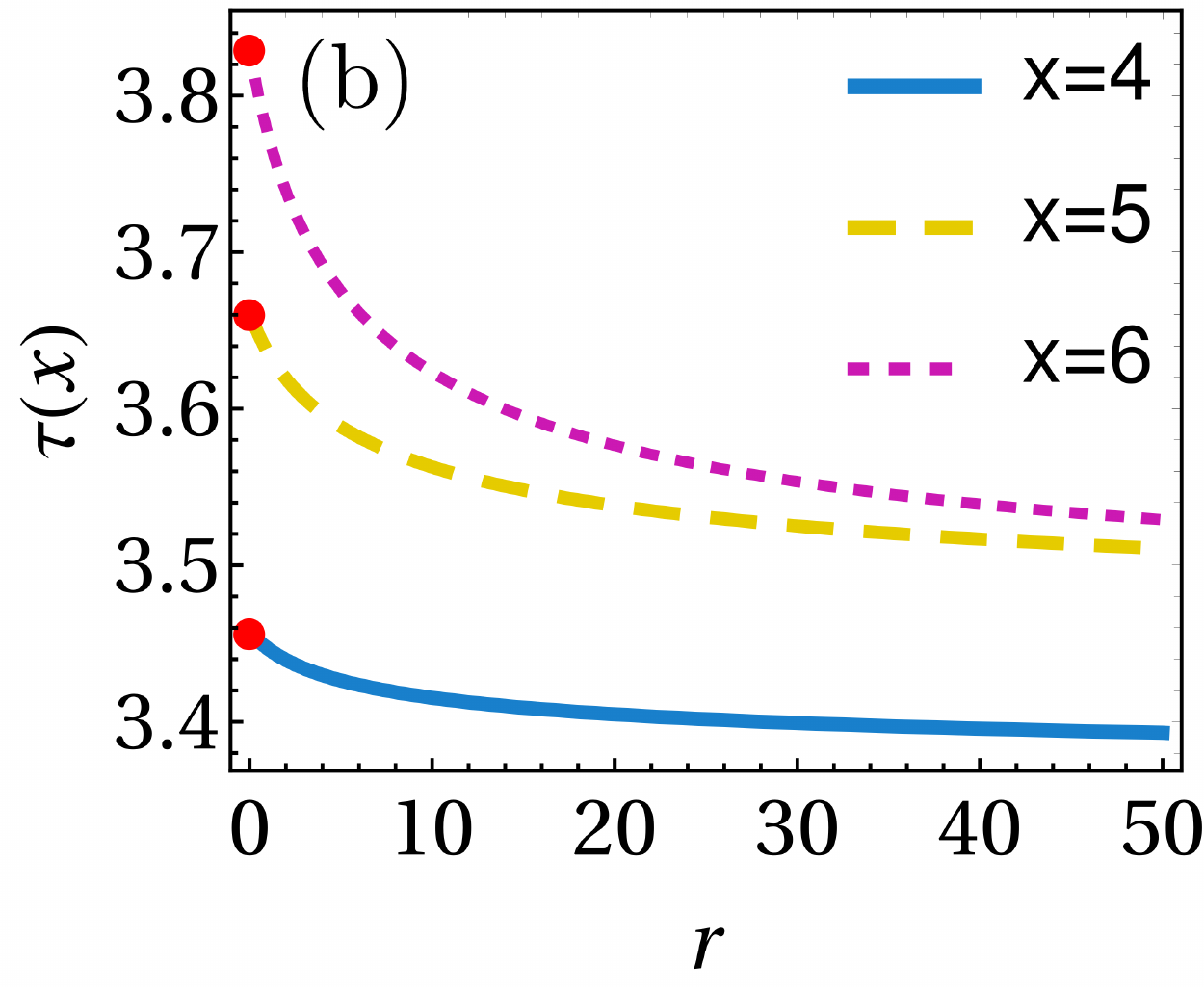}
    \caption{Mean first passage time  $\tau(x)$ [given in \eqref{mfpt-diff-process-m} and \eqref{mfpt-db-process-m}] with respect to resetting rate $r$ for given $x$. Herein, we show $\tau(x)$ for diffusion model in panel (a) and the ecological model in panel (b). As it is reasonable, the mean first passage time decreases with resetting rate $r$ for given $x$. For $r\to 0$, $\tau(x)$ diverges only for the diffusion model while it stays finite (indicated by filled circles in panel (b)) for the ecological setting, and is in agreement with the mean first passage time in the absence of resetting. The parameters for panel (a) are $D=1,~x_r=5$ and for panel (b) $b=0.5,~x_r=5$.}
    \label{fig:bd-bv}
  \end{center}
\end{figure}

In Fig.~\ref{fig:bd-bv}, the behavior of the mean first passage time $\tau(x)$ is shown for two different models: diffusion system (panel (a)) and ecological model (panel (b)). It is clear that the $\tau(x)$ is monotonically decreasing with the resetting rate $r$ for given $x$. This is because the (asymmetric) resetting always brings the system close to the absorbing location in stark contrast to the symmetric resetting where system can also reset to the opposite direction to the absorbing location leads to non-monotonic behavior as shown in the seminal work by Evans and Majumdar \cite{Evans_2011}.  

\section{Conclusions}
\label{seccons}
In this work, we have studied an asymmetric state-dependent resetting mechanism for diffusive processes on the positive real line. This model has allowed us to obtain both \textit{i)} the stationary state when the system is subject to reflecting boundary conditions and \textit{ii)} the mean first passage time to the the origin. We have exactly derived these quantities in detail for two different model systems: the paradigmatic homogeneous diffusion process, and an ecological model for species-rich ecosystems. In both cases, numerical simulations are in perfect agreement with our theoretical predictions, validating our results.

An important motivation to study this class of models with asymmetric resetting relies on ecological applications. We have modeled the effect of a catastrophic event as a sudden drop of the population to a fixed value $x_r>0$. Such extreme events, owing to environmental changes, may have disruptive consequences on ecosystems. This is of course a caricature of reality, but this toy model is nevertheless a good starting point that allows exact mathematical treatment and initial investigations of ecological or economic crashes. We have obtained that the MFPT, which is the average time for a species to become extinct, always decreases with the disaster rate $r$. This is an intuitive result that contrasts with the usual symmetric resetting in Brownian dynamics \cite{Evans_2011b}, where the optimal resetting rate can be derived. However, in our framework with asymmetry, the reset event always drives the system closer to the absorbing position, thus decreasing the first passage time on average. 

As well as developing new interesting theoretical aspects of non-equilibrium statistical mechanics, asymmetric stochastic resetting is an appealing tool for understanding fundamental features of natural disaster dynamics in different systems, including ecosystems. A good deal of realism could be achieved by considering $ x_r $ a quenched random variable. The final stationary distributions and the MFPT should be averaged over the probability density function of $ x_r $, thus increasing the variability of the final distributions.

The presented model is also applicable to other fields beyond ecology and statistical mechanics. For instance, the ecological model we have previously outlined is known as the Cox–Ingersoll–Ross model \cite{cox_1985} in the mathematical finance literature. Such a paradigmatic model with asymmetric resetting could be considered a first approximation when including the effects of sudden financial crises.\\

\appendix

\section{Ecological model in dimensionless variables}
\label{sec:app_dimensionless}

The ecological model used in our work was introduced and studied in detail in \cite{azaele2006}. This model stems from a continuous description of a birth and death process. Specifically, the drift and diffusion coefficients, respectively, are given by
\begin{equation}
\label{eq:coef-eco}
A(x)=b- \mu x, \quad B(x)=D x.
\end{equation} 
Herein, there are three biological parameters, namely, $\mu$,
$b$, and $D$. First, $\mu$ is the inverse of the characteristic time associated with species turnover. Second, $b$ takes into account the effects from immigration. Finally, $D$ accounts for the demographic stochasticity. 

It is handy to use a dimensionless description defined by  new variables $\tilde{x}=\mu x/D$, $\tilde{t}= \mu t$, and parameters $\tilde{b}=b/D$, $\tilde{r}=r/\mu$. Of course, the new timescale enters also in the definition of the mean first passage time, $\tilde{\tau}= \mu \tau$. For the sake of simplicity, in our notation we drop the tildes from now on. Using these dimensionless variables and parameters, we have the drift and diffusion terms:
\begin{equation}
\label{eq:coef-eco}
A(x)=b- x, \quad B(x)= x.
\end{equation} 
Remarkably, once we define proper scales the stochastic model without resetting reduces the number of parameters from three to one parameter.

\section{Explicit solution for the stationary state}
\label{sec:app_pss}
For the general case, the equation for the stationary distribution in presence of asymmetric resetting is the solution of the integro-differential equation,
\begin{align}
0 =  &-\partial_x[A(x)p_{ss}(x)] + \partial_x^2[B(x)p_{ss}(x)] \nonumber \\  &- r\Theta (x-x_r)p_{ss}(x) + r \delta (x-x_r) \int_{x_r}^{\infty} dy~ p_{ss}(y),\label{eqn1}
\end{align}
submitted to (i) natural boundary conditions in zero and infinity, and the matching conditions discussed in the main text, (ii) the matching condition at $x_r$, and (iii) the normalization from zero to infinity.

\subsection{Case (I): Pure diffusion}
First,  we consider the simplest homogeneous diffusive process, i.e., $B(x)=D$ in the absence of any drift $A(x)=0$. Hence, this is a pure diffusion process on the positive side of $x$-axis subjected to an asymmetric resetting mechanism. 
In this case, the probability flux is given by $-D \partial_x p_{ss}(x)$ [see Eq.~\eqref{eqn1}]. Therefore, the solutions to the left and to the right of $x_r$ can be computed. Specifically, we find that
\begin{subequations}
\begin{align}
f_L(\mathcal{N}_1,x)&=\mathcal{N}_1, \\
f_R(\mathcal{N}_2,x)&=\mathcal{N}_2~e^{-x \sqrt{r/D}},
\end{align}
\end{subequations}
where $\mathcal{N}_1$ and $\mathcal{N}_2$ are the constants that can be determined using the matching and normalization conditions discussed in the main text, and we obtain 
\begin{align}
\mathcal{N}_1&=\dfrac{1}{x_r+\sqrt{D/r}},\nonumber\\
\mathcal{N}_2&=\dfrac{ e^{x_r\sqrt{r/D}}}{x_r+\sqrt{D/r}}.\nonumber\\
\end{align}
Finally, substituting these $\mathcal{N}_{1,2}$, we find the stationary probability density function given in Eq.~\eqref{pss-I}.

\subsection{Case (II): Ecological model}
Now, we focus on solving the stationary distribution in the ecological model defined in \eqref{eq:coef-eco}. The solutions to the left and to the right of $x_r$ can be computed, and we get 
\begin{subequations}
\begin{align}
f_L(\mathcal{N}_1,x)&=\mathcal{N}_1 x^{-1+b}~e^{-x}, \\
f_R(\mathcal{N}_2,x)&= f_L(\mathcal{N}_2,x)~U(r,b,x),
\label{bd-eq-ss}
\end{align}
\end{subequations}
where $U(a,b,x)$ is the confluent hypergeometric function of the second kind also known as Tricomi's function. Imposing the matching and normalization conditions, we obtain value of constants $\mathcal{N}_1$ and $\mathcal{N}_2$ in terms of the parameter of the model,
\begin{subequations}
\begin{align}
\mathcal{N}_1=& \Gamma (1+r) \Gamma (1+r-b) U(r,b,x_r) 
\nonumber \\ 
& \times  \bigg[ x_r^b \Gamma(-b) \Gamma(1+r) \, {}_1F_1(b-r,1+b,-x_r) \nonumber \\
& \qquad + \Gamma(1+r-b) \left\{ \Gamma(b) \, {}_1F_1(-r,1-b,x_r) \right. \nonumber \\ & \qquad +\left. \Gamma(1+r) U(r,b,x_r) \left[\Gamma(b)-\Gamma(b,x_r)\right] \right\} \bigg]^{-1}, \\
\mathcal{N}_2=& \frac{\mathcal{N}_1}{U(r,b,x_r)},
\end{align}
\end{subequations}
where $_1F_1(\alpha;\beta;x)$ is the Kummer confluent hypergeometric function, and $\Gamma(z):=\int_0^{\infty}dt~e^{-t}t^{z-1}$ and $\Gamma(z,a):=\int_a^{\infty}dt~e^{-t}t^{z-1}$, respectively, are the gamma and the incomplete gamma functions. Thus, we obtain the final distribution as shown in Eq.~\eqref{pss-II}, where for simplicity, we write $\mathcal{N}=\mathcal{N}_1$.

\section{Derivation of the equation for the mean first passage time}
\label{sec:app_eqMFPT}
In this section, we obtain the mean first passage time for the system to hit the target $x=0$ (i.e., the absorbing boundary) for the first time during the evolution. It is always convenient to write the backward master equation. With this, we study  the probability density function $p(x,t|x_0,t_0)$ for the system  to be in $x$ at time $t$ starting from $x_0$ at time $t_0$ as a function of $x_0$ and $t_0$. Note that in the backward equation, both $x_0$ and $t_0$ are the variables in contrast to the case of forward formalism where they play the role of parameters with $x$ and $t$ being the variables. Our starting point to derive the backward framework is the Chapman-Kolmogorov equation \cite{vK},
\begin{equation}
p(x,t|x_0,t_0)=\int_0^\infty dx_1~ p(x,t|x_1,t_1) p(x_1,t_1|x_0,t_0),\label{ck-eqn}
\end{equation}
where $t_1\in (t,t_0)$ is an intermediate time. If we differentiate the above Eq. \eqref{ck-eqn} with respect to $t_1$, introduce the forward equation for $p(x_1,t_1|x_0,t_0)$, carry out integration by parts and evaluate it at the end for $t_1=t_0$, we finally arrive at
\begin{align}
-\dfrac{\partial p(x,t|x_0,t_0)}{\partial t_0}=&\bigg[A(x_0)\dfrac{\partial}{\partial {x_0}}+B(x_0)\dfrac{\partial^2}{\partial {x_0^2}}\bigg] p(x,t|x_0,t_0) \nonumber \\ &+r \Theta(x-x_r) \nonumber \\ & \quad \times \big[p(x,t|x_r,t_0)-p(x,t|x_0,t_0) \big].
 \label{BFP}
\end{align}
The above equation is the desired backward master equation.

Integrating the above equation \eqref{BFP} over $x$ from $0$ to $\infty$, shifting  $t_0$ by changing the variable $t-t_0$ to $t$, and then, differentiating with respect to time $t$, we obtain the evolution equation for the first passage distribution $F(t,x)$ for a system departing from $x$ and arriving at $x=0$ for the first time,
\begin{align}
\label{eq:BFP}
 \dfrac{\partial F(t,x)}{\partial t}=&\bigg[A(x)\dfrac{\partial}{\partial {x}}+B(x)\dfrac{\partial^2}{\partial {x_0^2}}\bigg] F(t,x)\nonumber \\ &+r \Theta(x-x_r) \big[F(t,x_r)-F(t,x) \big].
\end{align}
Note that in order to simplify the notation we have dropped the subindex $0$ in $x_0$.
The above equation is subjected to the boundary conditions $F(0,x)=0$ and $\lim_{t\to\infty} F(t,x)=0$, where the latter condition ensures that $\int_0^\infty dt~ F(t,x)$ is finite. 

Now, we define the probability of exiting through $x=0$ departing from $x$ regardless of the time required 
\begin{equation}
\Pi(x):= \int_0^\infty dt~F(t,x),
\end{equation}
where the boundary conditions for $\Pi(x)$ are
\begin{subequations}
\label{eq:pi-bc}
\begin{align}
\label{eq:pi-bc1}
&\Pi(0)=1,
\\
\label{eq:pi-bc2}
\lim_{x \to \infty} &\Pi(x)  \text{ is finite}.
\end{align}
\end{subequations}
While the first condition ensures the total exit probability of the system started from the absorbing boundary is one, the second one says that there is a finite probability of the system to reach the absorbing boundary at $x=0$ started from $x\to\infty$.  

This quantity follows the following differential equation
\begin{align}
\label{eq:pi-ev}
0 =&  A(x)\partial_{x}\Pi(x)+B(x) \partial_x^2 \Pi(x) \nonumber \\ & +r \Theta(x-x_r) \left[\Pi(x_r)-\Pi(x) \right].
\end{align}
The solution of Eq. \eqref{eq:pi-ev} given the boundary conditions \eqref{eq:pi-bc} for the two cases of interest we have already introduced in this paper is simply $\Pi(x)=1$ since the system eventually reach the absorbing boundary.

Now, the mean first passage time $\tau(x)$ for exiting through $x=0$ is defined as 
\begin{equation}
\tau(x):=\dfrac{\int_0^\infty dt \,\,t\, F(t,x)}{\Pi(x)}. 
\label{mean-time-def}
\end{equation}
Multiplying equation \eqref{eq:BFP} by $t$ and integrating over time from $0$ to $\infty$,  we obtain the differential equation for $\tau(x)$,
\begin{align}
\label{eq:tau-ev}
-\Pi(x) =&  A(x) \partial_{x}[\Pi(x)\tau(x)]+B(x) \partial_x^2 [\Pi(x)\tau(x)]  \nonumber \\ &+r \Theta(x-x_r) \left[\Pi(x_r)\tau(x_r)-\Pi(x)\tau(x) \right],
\end{align}
where we have made use of $\lim_{t \to \infty} t F(t,x)=0$.
The boundaries condition in this case are
\begin{subequations}
\label{eq:tau-bc}
\begin{align}
\label{eq:tau-bc1}
&\tau(0)=0,
\\
\label{eq:tau-bc2}
\lim_{x \to \infty} &\tau(x) ~ \text{is  finite}.
\end{align}
\end{subequations}
Note that the presence of resetting provides that the mean first passage time has to be finite for $x \to \infty $ since the reset connects any value of $x>x_r$ with $x_r$.

Eqs. \eqref{eq:pi-ev} and \eqref{eq:tau-ev} can be solved to the left and to the right of $x_r$ separately. Boundary conditions \eqref{eq:pi-bc1} and \eqref{eq:tau-bc1} apply to the left solution whereas the   \eqref{eq:pi-bc2} and \eqref{eq:tau-bc2} apply to the right solution. The full solution of the both $\Pi(x)$ and $\tau(x)$ can be obtained using matching condition at $x_r$ (i.e., both functions and their first derivatives should be continuous at $x=x_r$). However, these are difficult to obtain for general drift and diffusive coefficient. In the following section, we study in detail the two cases of interest taking into account that $\Pi(x)=1$ therein.

\section{Explicit solution for the mean first passage time}
\label{sec:app_MFPT}
The equation for the mean first passage time in the general case is given by \eqref{eq:tau-ev} submitted to boundary conditions in \eqref{eq:tau-bc} and the matching condition. Below, we study the two cases of interest reported in the main text.
\subsection{Case (I): Pure diffusion}
In the case of pure diffusion, the differential equation for $\tau(x)$ becomes simply
\begin{equation}
-1 = D \partial_x^2 \tau(x) + r \Theta(x-x_r) [\tau(x_r) - \tau(x)].
\end{equation}
We solve the above differential equation using the boundary conditions \eqref{eq:tau-bc} and matching conditions and get the solution reported in the main text
\begin{align}
\label{mfpt-diff-process}
\tau(x)=\begin{cases}
- \dfrac{x^2}{2D} + x\left(\dfrac{x_r}{D}+ \dfrac{1}{\sqrt{rD}} \right) & 0\leq x \leq x_r,\\
\\
\dfrac{1-e^{-(x-x_r)\sqrt{r/D}}}{r}+\dfrac{x_r^2}{2D} + \dfrac{x_r}{\sqrt{rD}}& x > x_r.
\end{cases}
\end{align}
\subsection{Case (II): Ecological model}
In the ecological case, we find again that $\Pi(x)=1$. 
Thus, the mean first passage time $\tau(x)$ obeys the differential equation
\begin{align}
-1 =& (b- x ) \partial_x \tau(x)+  x \partial_x^2 \tau(x) \nonumber \\&+ r \Theta(x-x_r) [\tau(x_r) - \tau(x)].
\label{mft-ec}
\end{align}
It is possible to solve the above differential equation using the boundary conditions \eqref{eq:tau-bc} and the matching conditions at $x=x_r$. That yields the solution
\begin{equation}
\label{mfpt-db-process}
\tau(x)=\tau_L(x) \Theta(x_r-x)+\tau_R(x)\Theta(x-x_r),
\end{equation}
where
\begin{subequations}
\begin{align}
\tau_L(x)= & \int_0^x  \, dy \, y^{-b}e^y \bigg[ \Gamma(b,y) - \Gamma (b,x_r) 
\nonumber \\& \qquad \qquad \qquad + \frac{U(1+r,1+b,x_r)}{U(r,b,x_r)} x_r^b e^{-x_r} \bigg] \label{eq:tauL} \\
\tau_R(x)= &\tau_L(x_r) + \frac{1}{r} \left[ 1- \frac{U(r,b,x)}{U(r,b,x_r)}\right],
\end{align}
\end{subequations}
which is the solution reported in the main text. 
The integral in Eq.~\eqref{eq:tauL} can be explicitly carried out. Nevertheless, we have chosen to keep the integral form in order to avoid clutter. Note that the above solutions is well defined for $b<1$, as also happened   in absence of resetting for the absorbing solution in the original model \cite{azaele2006}.

\section{Simulation method}
Herein, we put forward the method of numerical simulation we have used along this work. Specifically, all simulations are based on the the discretization of the  Langevin equation \eqref{dyn-1} which is complemented with the stochastic resetting.

To obtain the distribution at time $t$, we discretize the time $t=n~\Delta t$, where $n$ is an integer and $\Delta t$ stands for the unit time step. Our choice for the initial condition is $x(0)=x_r$. On the one hand, if $x(t)>x_r$,
\begin{enumerate}
    \item with probability $1-r \Delta t$, where $r$ is a constant resetting rate, the system evolves according to
\begin{align}
    x(t+\Delta t)=x(t) + A(x(t)) + \sqrt{2B(x(t))\Delta t}~\zeta,
\end{align}
where $\zeta$ is the Gaussian random variable with mean 0 and variance 1. 
\item whereas it is abruptly reset to $x_r$ with probability $r \Delta t$, \end{enumerate}
On the other hand, if $x(t)<x_r$, the system undergoes the stochastic evolution as illustrated in the step 1. The process iterates until time $t$ is achieved.
We obtain the distribution building the histogram after repeating the stochastic process $\mathcal{N}_R$ realizations.

For the results of the MFPT, we assume an absorbing boundary at the origin $x=0$, and observe the first time the system hits the absorbing boundary following discretized scheme of the dynamics as illustrated above. We repeat the process for $\mathcal{N}_R$ number of realizations and compute the MFPT.

\begin{acknowledgments}
C. A. P. acknowledges the support from University of Padova through Project No. STARS-Stg (CdA Rep. 40, 23.02.2018) BioReACT grant. D. G. is supported by ``Excellence Project 2018'' of the Cariparo foundation. We thank Amos Maritan and Samir Suweis for useful discussions.
\end{acknowledgments}


\end{document}